\documentclass[aps,prl,reprint,preprintnumbers,superscriptaddress,nofootinbib,floatfix]{revtex4-2}

\usepackage{amsmath,amssymb,amsfonts,bm}
\usepackage{graphicx}
\usepackage{xcolor}
\usepackage[normalem]{ulem}
\usepackage{booktabs}
\usepackage[colorlinks=true,linkcolor=blue,citecolor=blue,urlcolor=blue]{hyperref}
\begin{document}
\preprint{IPPP/26/91}

\newcommand{\UNISA}{\affiliation{Dipartimento di Fisica ``E.R.\ Caianiello'', Universit\`a degli Studi di Salerno, Via Giovanni Paolo II 132, 84084 Fisciano (SA), Italy}}
\newcommand{\INFNSA}{\affiliation{Istituto Nazionale di Fisica Nucleare, Gruppo Collegato di Salerno, Sezione di Napoli, Via Giovanni Paolo II 132, 84084 Fisciano (SA), Italy}}
\newcommand{\LNL}{\affiliation{Istituto Nazionale di Fisica Nucleare, Laboratori Nazionali di Legnaro, Viale dell'Universit\`a 2, I-35020 Legnaro (PD), Italy}}

\title{A proposed QUAX search for Lorentz violating scalar interactions}

\author{Domenico D'Agostino}
\email{domenico.dagostino@na.infn.it}
\INFNSA

\author{Giovanni Carugno}
\email{giovanni.carugno@pd.infn.it}
\LNL

\author{Gerardo Iannone}
\email{iannone@sa.infn.it}
\INFNSA

\author{Gaetano Lambiase}
\email{glambiase@unisa.it}
\UNISA
\INFNSA

\author{Antonello Ortolan}
\email{ortolan@lnl.infn.it}
\LNL

\author{Tanmay Kumar Poddar}
\email{tanmay.k.poddar@durham.ac.uk}
\affiliation{Institute for Particle Physics Phenomenology (IPPP), Department of Physics, Durham University, Durham DH1 3LE, United Kingdom}

\author{Giuseppe Ruoso}
\email{giuseppe.ruoso@lnl.infn.it}
\LNL

\author{Luca Visinelli}
\email{lvisinelli@unisa.it \newline }
\UNISA
\INFNSA

\begin{abstract}
We propose a dedicated QUAX-$g_pg_s$ measurement of Lorentz violating scalar mediated spin interactions using a modulated laboratory nucleon source and a paramagnetic electron spin detector. We calculate the finite source and detector response to an electron coupling with a fixed direction in the Sun centered celestial equatorial frame. The source modulation, combined with the sidereal evolution of the detector projection, provides sensitivity to the mixed coupling $g_N\widetilde{\mathbf J}^{e}$. For an east pointing detector, the two equatorial components are extracted through a simultaneous fit to the signed source referenced quadratures. We derive the corresponding pseudo magnetic field and induced magnetization and specify the readout, timing, and background assumptions entering the forecast. For $11\,\mathrm{h}$ of live time distributed over $72\,\mathrm{h}$, the projected individual 95\% intervals from the two component fit reach approximately $1.5\times10^{-43}$ at low mediator mass for a magnetization equivalent noise of $10\,\mathrm{aT}/\sqrt{\mathrm{Hz}}$. This result defines a conditional benchmark for a local source search.
\end{abstract}

\maketitle
\raggedbottom

\textbf{\textit{Introduction. ---}}
The Standard Model (SM) of particle physics and General Relativity are built upon Lorentz and charge conjugation-parity-time reversal (CPT) invariance, and precision tests of these symmetries provide sensitive probes of new physics~\cite{Colladay:1998fq, Mattingly:2005re, Liberati:2013xla, Kostelecky:2008ts}. Possible departures from Lorentz invariance can be parametrized through background vector and tensor coefficients, conventionally taken to be constant in an approximately inertial Sun-centered frame. While spin-dependent interactions beyond the SM have been extensively investigated~\cite{Moody:1984ba, Dobrescu:2006au, Cong:2024qly, Arza:2026rsl}, scalar-mediated Lorentz- and CPT-violating interactions remain comparatively less explored~\cite{Altschul:2012xu} and constitute the focus of this work.

\begin{figure}[tbh]
	\centering
	\includegraphics[width=0.75\linewidth]{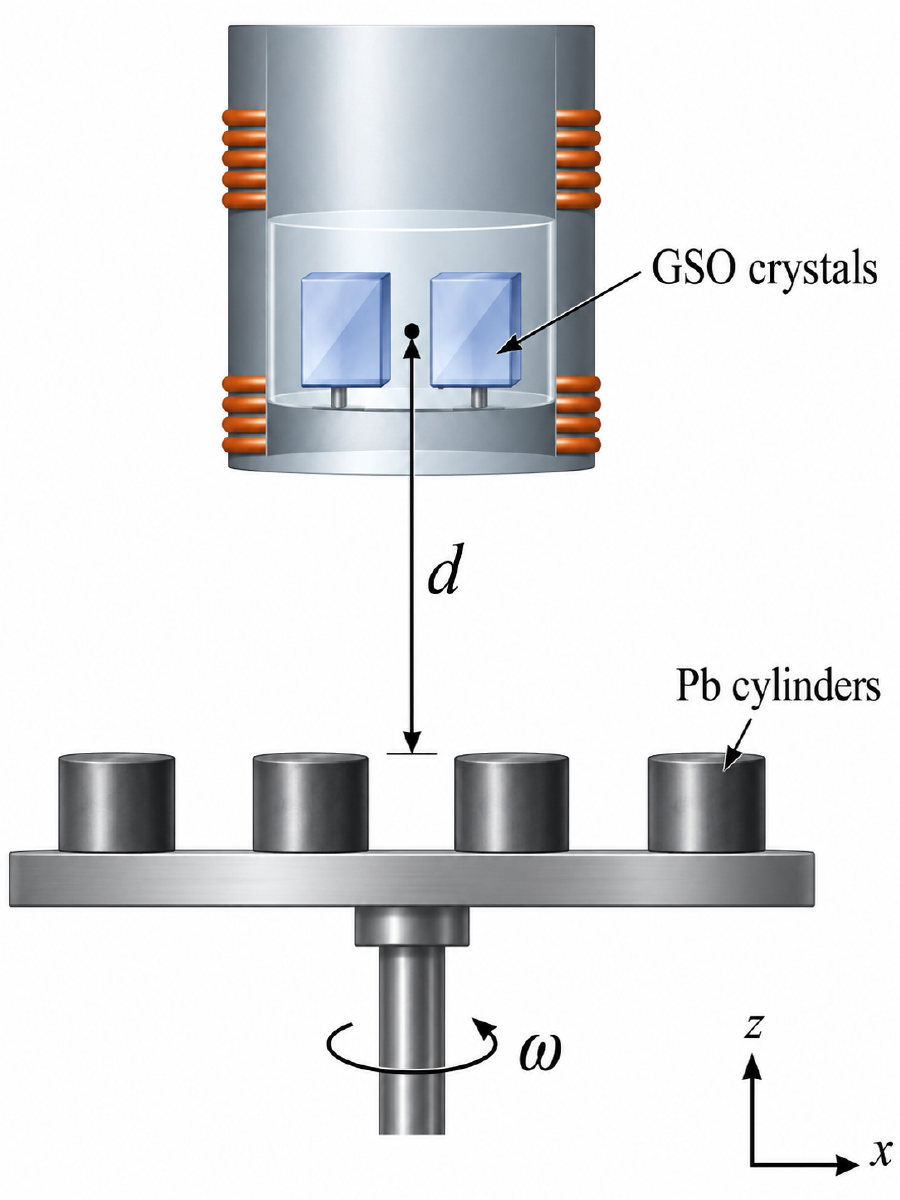}
    \caption{Schematic of the QUAX-LV setup. A rotating lead source modulates the scalar field, inducing a magnetization in two paramagnetic GSO crystals measured by superconducting pickup coils coupled to a single SQUID. The source consists of two radial tracks on a horizontal wheel.}
	\label{fig:source_schematic}
\end{figure}

Existing torsion balance and magnetometer measurements have been used to derive stringent constraints on scalar-mediated spin-dependent interactions involving Lorentz and CPT violation in the ultralight scalar mass regime, $m_\phi\ll10^{-11}~\mathrm{eV}$, by exploiting astronomical bodies as macroscopic sources of the scalar field~\cite{Heckel:2008hw, Brown:2010dt, Smiciklas:2011xq, Tsai:2021irw, Tsai:2023zza, Carenza:2025jwn}. By contrast, controlled laboratory sources provide a complementary approach, allowing precise control over the source geometry, position, and modulation while probing mediator masses corresponding to interaction ranges characteristic of laboratory length scales~\cite{Crescini:2016lwj,Crescini:2017uxs,Crescini:2020ykp}.

In this {\it Letter}, we propose a dedicated search for scalar-mediated Lorentz-violating (LV) interactions with the QUest for AXions experiment~\cite{Crescini:2016lwj}, using a periodically modulated nucleon source and a paramagnetic electron spin detector, as illustrated in Fig.~\ref{fig:source_schematic}. This approach, here QUAX-LV, is complementary to the search for axion-mediated monopole-dipole interactions conducted by QUAX22~\cite{Crescini:2020ykp}. The QUAX22 apparatus employed rotating lead masses at room temperature and a paramagnetic $\mathrm{Gd_2SiO_5}$ (GSO) detector operated at $4.2\,\mathrm{K}$, with the induced magnetization measured by a superconducting quantum interference device (SQUID). The experiment reached a magnetic field noise of $53\,\mathrm{aT}/\sqrt{\mathrm{Hz}}$ near 6\,Hz and collected approximately 11\,h of data~\cite{Crescini:2020ykp}. These parameters provide the experimental reference for the operating frequency and live time adopted in our sensitivity forecast, while the source and detector geometries are optimized for the Lorentz-violating search.

Unlike the conventional scalar-pseudoscalar monopole-dipole interaction, the scalar considered here couples to nucleons through $g_N$ and to electron spins through the Lorentz-violating vector $\widetilde{\mathbf J}^{e}$, whose direction is fixed in the Sun-centered celestial equatorial frame (SCF). Its laboratory projection varies with sidereal time~\cite{Kostelecky:1999mr, Kostelecky:2008ts, Heckel:2008hw, Brown:2010dt, Smiciklas:2011xq}. The controlled source modulation and the Earth's rotation produce distinct time dependences that can be analyzed directly without resolving separate sidereal sidebands. While the underlying interaction and its magnetometer interpretation are established~\cite{Altschul:2012xu,Carenza:2025jwn}, we develop the finite geometry response and a directional acquisition and inference framework for probing the mixed coupling $g_N\widetilde{\mathbf J}^{e}$. Complementary proposals involving Josephson junctions~\cite{Croon:2026jj}, hydrogen masers, and molecular spin precession~\cite{Croon:2026fvx} have also been investigated. We use natural units with $\hbar=c=k_B=1$.

\begin{figure}[tbh]
	\centering
	\includegraphics[width=0.8\linewidth]{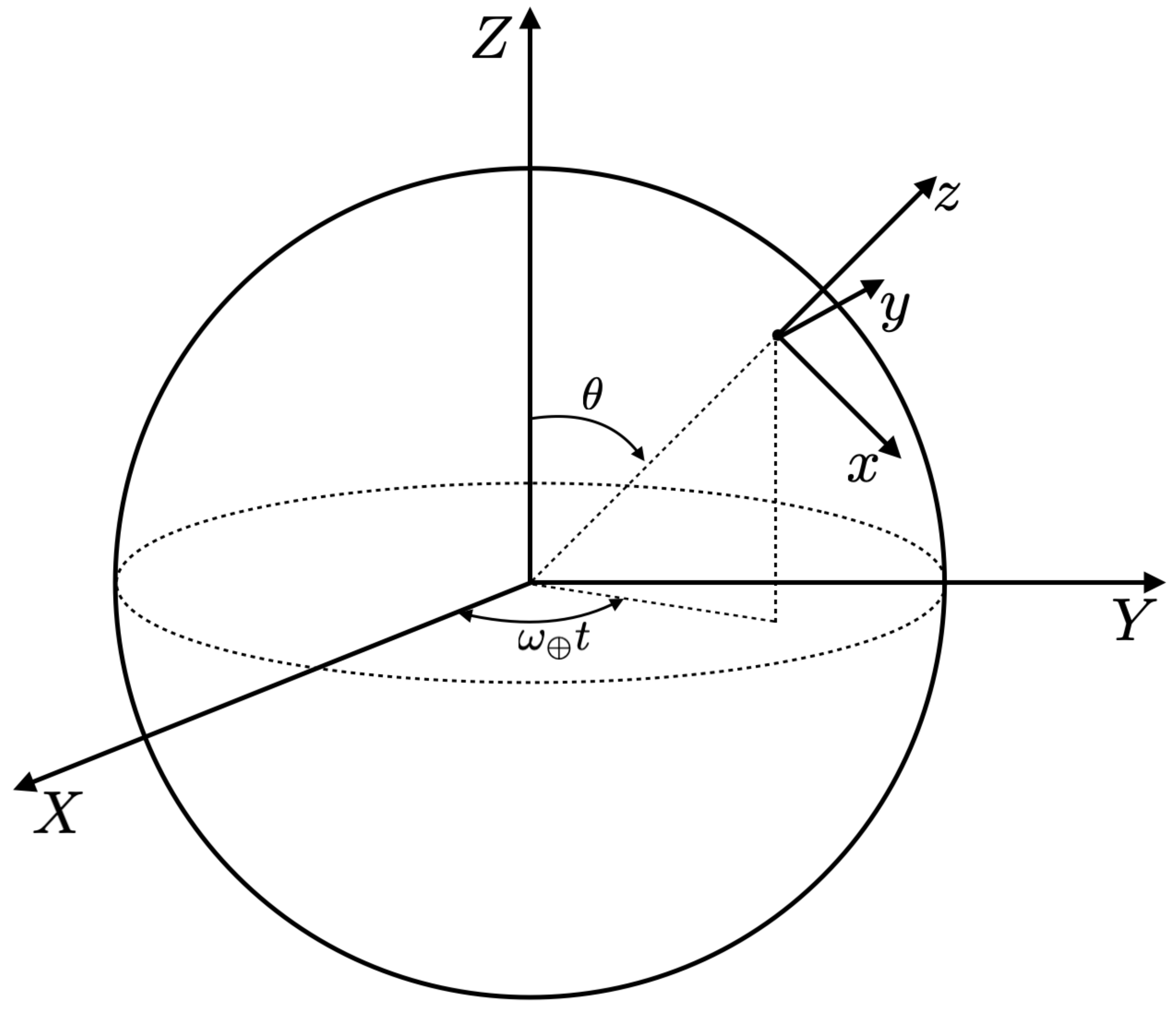}
	\caption{Schematic relation between the SCF with axes $(X,Y,Z)$ and the local laboratory frame with axes $(x,y,z)$. The laboratory frame rotates with the Earth at angular frequency $\omega_\oplus$, and the site colatitude is denoted by $\theta$.}
	\label{fig:geometry}
\end{figure}

\textbf{\textit{Lorentz violating scalar interactions. ---}}
We consider a light scalar boson $\phi$ coupled to nucleons and electrons through the interaction Lagrangian~\cite{Altschul:2012xu,Carenza:2025jwn}
\begin{equation}
	\mathcal{L}_{\rm int}\supset-g_N\phi\bar NN-J^e_\mu \, \phi\, \bar e\gamma_5\gamma^\mu e-\frac{1}{2}\phi L^e_{\mu\nu} \, \bar e\sigma^{\mu\nu}e\,,
	\label{eq:Lint}
\end{equation}
where $\sigma^{\mu\nu}=i[\gamma^\mu,\gamma^\nu]/2$. Here, $N$ and $e$ are the nucleon and electron Dirac fields, respectively, while $J^e_\mu$ and $L^e_{\mu\nu}$ are constant Lorentz violating coefficients in the electron sector. The scalar coupling to nucleons, $g_N$, is assumed to weight protons and neutrons equally, $g_N=g_p=g_n$, so that an unpolarized source can be described by its nucleon number density. We neglect terms suppressed by the laboratory velocity and retain the leading spin interaction in the nonrelativistic limit. In the SCF, we define the effective spatial coefficient
\begin{equation}
	\widetilde J^e_A \equiv J^e_A + \epsilon_{ABC}L^e_{BC}\,, \qquad A=X,Y,Z\,,
	\label{eq:Jtilde}
\end{equation}
with $\epsilon_{XYZ}=+1$. The corresponding laboratory components are
\begin{equation}
    \widetilde J^e_{{\rm lab},j}(t) \equiv R_{jA}(t)\widetilde J^e_A\,,
	\label{eq:Jlab}
\end{equation}
where $R_{jA}(t)$ rotates the SCF components into the local laboratory basis. The geometry of the two frames is shown in Fig.~\ref{fig:geometry}. The SCF is the right-handed basis $(X,Y,Z)$, with the $Z$ axis aligned with the Earth's rotation axis and the $X$ axis pointing toward the vernal equinox. At the laboratory position, the local basis $(x,y,z)$ is chosen with $z$ along the local vertical, $y$ pointing east, and $x$ pointing south. Setting the colatitude $\theta$ and the sidereal angular frequency $\omega_\oplus$, Eq.~\eqref{eq:Jlab} gives
\begin{align}
	\widetilde J^e_{{\rm lab},x}(t)&=\cos\theta\left(\widetilde J^e_X\cos\omega_\oplus t \!+\! \widetilde J^e_Y\sin\omega_\oplus t\right) \!-\! \widetilde J^e_Z\sin\theta, \label{eq:Jxlab}\\
	\widetilde J^e_{{\rm lab},y}(t)&=-\widetilde J^e_X\sin\omega_\oplus t+\widetilde J^e_Y\cos\omega_\oplus t, \label{eq:Jylab}\\
	\widetilde J^e_{{\rm lab},z}(t)&=\sin\theta\left(\widetilde J^e_X\cos\omega_\oplus t \!+\! \widetilde J^e_Y\sin\omega_\oplus t\right) \!+\! \widetilde J^e_Z\cos\theta. \label{eq:Jzlab}
\end{align}
For the QUAX-LV benchmark, we consider a detector configuration in which the relevant sensitivity axis is aligned with the local eastward direction $\hat y$. Consequently, the accessible Lorentz-violating component is $\widetilde J_{\mathrm{lab},y}^{e}(t)$, which depends only on the equatorial Sun-centered coefficients $\widetilde J_X^e$ and $\widetilde J_Y^e$, as shown in Eq.~\eqref{eq:Jylab}.

For a source with nucleon number density $n_N(\mathbf x',t)$, the scalar field obeys $(\nabla^2-m_\phi^2)\phi = g_N n_N$, with solution vanishing at infinity
\begin{equation}
	\phi(\mathbf x,t) \equiv -g_N\Phi(\mathbf x,t) = -g_N\int_{V_s(t)}\!\!{\rm d}^3{\bf x'}\frac{n_N(\mathbf x',t)}{4\pi}\frac{e^{-m_\phi|\mathbf x-\mathbf x'|}}{|\mathbf x-\mathbf x'|}\,,
    \label{eq:phisource}
\end{equation}
where $V_s(t)$ is the instantaneous source volume and $\Phi(\mathbf x,t)$ is the coupling-independent source function. The spin Hamiltonian generated by this field is
\begin{equation}
	\mathcal H_{\rm spin}(\mathbf x,t) = \phi(\mathbf x,t)\,\sigma_e^j\, \widetilde J^e_{{\rm lab},j}(t) = \phi(\mathbf x,t)\, \bm{\sigma}_e\cdot \widetilde{\mathbf J}^{e}_{\rm lab}(t)\,.
	\label{eq:Hspin}
\end{equation}
The electron spin interaction depends only on the combination $\widetilde{\mathbf J}^{e}$ and does not separately determine the $J^e_A$ and $L^e_{BC}$ contributions. Comparing Eq.~\eqref{eq:Hspin} with the magnetic interaction $\mathcal H_B=-\mu_e\bm{\sigma}_e\cdot\mathbf B$, where $\mu_e$ is the electron magnetic moment, we define the equivalent pseudo magnetic field
\begin{equation}
    \mathbf B^\phi_p(\mathbf x,t) = -\frac{\phi(\mathbf x,t)}{\mu_e}\widetilde{\mathbf J}^{e}_{\rm lab}(t)\,.
    \label{eq:Bpseudo}
\end{equation}
For the east pointing QUAX-LV configuration, Eqs.~\eqref{eq:phisource} and~\eqref{eq:Jylab} give
\begin{equation}
	B^\phi_{p,y}(\mathbf x,t) = \frac{g_N\Phi(\mathbf x,t)}{\mu_e} \left[-\widetilde J^e_X\sin\omega_\oplus t +\widetilde J^e_Y\cos\omega_\oplus t \right].
	\label{eq:Bpy}
\end{equation}
Equivalently, the corresponding spin-dependent potential for a single electron is
\begin{equation}
	V_{\rm LV}(\mathbf x,t)=-g_N\Phi(\mathbf x,t)\,\bm{\sigma}_e\cdot\widetilde{\mathbf J}^{e}_{\rm lab}(t)\,.
	\label{eq:Vsource}
\end{equation}
This potential is distinct from the monopole--dipole interaction searched for in QUAX22. For a pseudoscalar electron coupling and a scalar nucleon coupling, the nonrelativistic interaction contains the radial direction $\hat{\mathbf r}$ associated with each source-detector separation, so that its spin dependence depends on the position. In Eq.~\eqref{eq:Vsource}, by contrast, the underlying Lorentz violating coefficient $\widetilde{\mathbf J}^{e}$ is fixed in the SCF, and its laboratory components vary only through the rotation of the Earth.

\textbf{\textit{Experimental configuration. ---}}
The experimental starting point for QUAX-LV is the QUAX22 spin dependent fifth force apparatus. The detector consisted of two paramagnetic GSO crystals with dimensions $2.5\times2.5\times1.1\,\mathrm{cm}^3$, coupled through parallel superconducting pickup coils to one SQUID magnetometer. The source consisted of $n_s=12$ pairs of lead cylinders mounted on a horizontal aluminium disk of about 1\,m diameter. The minimum distance between the GSO center of mass and the surface of a lead cylinder was approximately $3.5\,\mathrm{cm}$. A wheel rotation frequency $f_w\simeq0.5\,\mathrm{Hz}$ produced a source modulation at $f_s=n_s f_w \simeq 6\,\mathrm{Hz}$, where a magnetic field noise of $53\,\mathrm{aT}/\sqrt{\mathrm{Hz}}$ was measured and the noise followed the expected white noise scaling over approximately $4.1\times10^4\,\mathrm{s}$~\cite{Crescini:2020ykp}.

For a configuration of 24 lead cylinders with uniform density, the source function in Eq.~\eqref{eq:phisource} becomes
\begin{equation}
	\Phi(\mathbf x,t)=\frac{n_N}{4\pi}\sum_{a=1}^{24}\int_{V_a(t)}{\rm d}^3{\bf x'}\,\frac{e^{-m_\phi|\mathbf x-\mathbf x'|}}{|\mathbf x-\mathbf x'|}\,,
	\label{eq:PhiQUAX}
\end{equation}
where $V_a(t)$ is the spatial region occupied by lead cylinder $a$, so that the total source region is $V_s=\sum_a V_a$. The source is quasistatic on the light crossing time of the apparatus. In the linear response regime and with the magnetic field component in Eq.~\eqref{eq:Bpy}, the east component of the pseudo magnetic field induces a magnetization
\begin{equation}
	\mu_0 M^\phi(\omega)=\chi\,B^\phi_{p,y}(\omega)\,,
	\label{eq:magresponse}
\end{equation}
where we take the GSO magnetic susceptibility to be $\chi=0.7$ as in QUAX22 and neglect its frequency dependence over the narrow interval around the source modulation frequency. In QUAX22 the paramagnetic response was modeled as $\mu_0M=\chi B_p$, while the conversions from magnetization to pickup flux and from SQUID flux to voltage were calibrated separately. Here, the calibrated detector noise is referred to as an equivalent fluctuation of $\mu_0M$ in the GSO sample.

We describe the finite volume response by $\mathcal G(\Theta_w)$, where $\Theta_w$ is the physical wheel angle. Because of the rotational symmetry in the source track, $\mathcal G$ is periodic under $\Theta_w\to\Theta_w+2\pi/12$, corresponding to the source modulation frequency $f_s=12f_w\simeq6\,\mathrm{Hz}$. In the numerical benchmark, the superconducting pickup is assumed to weight the magnetization uniformly within each GSO crystal. The response of crystal $c=1,2$ is
\begin{equation}
	\mathcal G^{(c)}(\Theta_w)=\frac{\chi}{V_{\rm GSO}}\int_{V_c}{\rm d}^3{\bf x}\,\Phi(\mathbf x,\Theta_w)\,,
	\label{eq:volume_response}
\end{equation}
where $V_c$ is the spatial region occupied by crystal $c$ and $V_{\rm GSO} \equiv \int_{V_c}{\rm d}^3{\bf x} =2.5\times2.5\times1.1\,\mathrm{cm}^3$ is the volume of either crystal. For the benchmark, we assume that the parallel pickup and single SQUID readout weight the two crystals equally, so that
\begin{equation}
    \mathcal G(\Theta_w) = \frac{\mathcal G^{(1)}(\Theta_w)+\mathcal G^{(2)}(\Theta_w)}{2}\,. 
    \label{eq:Gavg}
\end{equation}
The part relevant to the search is the source modulated component
\begin{equation}
	\delta\mathcal G(t)=\mathcal G(\Theta_w(t))-\left\langle\mathcal G \right\rangle\,,
	\label{eq:dG}
\end{equation}
where brackets denote an average over one source modulation period. For a periodic source, we define the complex first harmonic response per unit mixed coupling over $T_s=1/f_s$ as
\begin{equation}
	\widetilde{\mathcal R}_1 = \frac{2}{T_s\,\mu_e}\int_0^{T_s}{\rm d}t\,\delta\mathcal G(t)e^{-i\omega_s t}\,,
	\label{eq:G1complex}
\end{equation}
with $\mathcal R_1\equiv|\widetilde{\mathcal R}_1|$. The normalization in Eq.~\eqref{eq:G1complex} gives the peak amplitude of a sinusoidal response, while the complex phase is kept in the source fit. All subsequent optimization and sensitivity calculations use this single combined readout.

As a cross-check of the finite geometry calculation, we repeated the response calculation using the monopole--dipole interaction searched for in QUAX22. The reconstructed response reproduces the published interaction range dependence to good accuracy up to an approximately constant normalization $\kappa_{2022}\simeq 4.24$. The comparison and its limitation are described in the Supplemental Material Sec.~\ref{app:geometry}.

Since QUAX-LV requires a new acquisition, the geometry need not reproduce the configuration of QUAX22. The source consists of two radial tracks, each containing 12 equally spaced lead cylinders, with one GSO crystal centered directly above each track and its $1.1\,\mathrm{cm}$ dimension oriented vertically. We vary the outer source track radius $R_{\rm out}$, the radial separation $\Delta R$ between the two tracks, the vertical distance $d$ from the GSO center of mass to the nearest Pb surface, and the relative azimuthal phase $\Delta\Theta$ between the tracks, with $R_{\rm in}=R_{\rm out}-\Delta R$. In this geometry, $d=2\,\mathrm{cm}$ corresponds to a minimum GSO surface to Pb surface gap of $1.45\,\mathrm{cm}$ and does not include the additional clearance required for the cryostat and shielding. For each trial geometry, the two crystal responses are combined as in Eq.~\eqref{eq:Gavg}, and the resulting complex first harmonic $\widetilde{\mathcal R}_1(m_\phi)$ enters the likelihood. The scan maximizes the mean of $\log\det F_{XY}$, with $F_{XY}$ the Fisher information matrix, over $m_\phi=(0.1,0.3,1,3,5,10,20)\,\mu\mathrm{eV}$, after profiling the constant source background. Table~\ref{tab:optimized_geometry} summarizes the scan ranges and the selected benchmark, with the detector sensitivity axis fixed to local east. Further optimization requires the measured pickup weighting and the relevant cryogenic, shielding, and mechanical constraints.

\begin{table}[thbp]
	\caption{Dedicated QUAX-LV benchmark selected from the geometry scan.}
	\label{tab:optimized_geometry}
	\begin{ruledtabular}
	\begin{tabular}{llc}
	Quantity & Scan range & Benchmark value \\
	\hline
	Outer source radius $R_{\rm out}$ & [38,45]\,cm & 45\,cm \\
	Track separation $\Delta R$ & [6,12]\,cm & 6\,cm \\
	Distance $d$ & [2,3.5]\,cm & 2\,cm \\
	Relative track phase $\Delta\Theta$ & $[0^\circ,15^\circ]$ & $0^\circ$
	\end{tabular}
	\end{ruledtabular}
\end{table}

\textbf{\textit{Acquisition and backgrounds. ---}}
An offset $\delta\Theta_w$ in the measured wheel angle produces a phase shift at the source modulation frequency by
\begin{equation}
	\varphi_s = n_s\,\delta\Theta_w+\varphi_{\rm geom}+\varphi_{\rm inst}(\omega_s)\,,
	\label{eq:phase_relation}
\end{equation}
where $\varphi_{\rm geom}$ is the phase of the calculated geometric response and $\varphi_{\rm inst}$ is the measured phase of the detector and electronics. As such, the source reference and SQUID data should be recorded with synchronized timestamps so that both the wheel angle and local sidereal time can be reconstructed. To suppress phase errors associated with fluctuations of the wheel angular velocity, the source phase can be reconstructed by recording interruptions of a laser beam by the rotating Pb masses with UTC timestamps from a GPS clock. The SQUID time stream can then be resampled in source phase before the in-phase and quadrature components are formed. The corresponding sidereal phase is reconstructed from the UTC timestamps and the laboratory coordinates and orientation.

We introduce the mixed couplings
\begin{equation}
	q_A\equiv g_N\widetilde J^e_A\,,\qquad A=X,Y,Z\,,
	\label{eq:qA}
\end{equation}
so that only $q_X$ and $q_Y$ enter the measurement for the east pointing configuration. The expected source modulated magnetization equivalent signal $s(t)$ follows from Eqs.~\eqref{eq:Bpy} and~\eqref{eq:dG} as
\begin{equation}
	s(t)=\frac{\delta\mathcal G(t)}{\mu_e}\left[-q_X\sin\omega_\oplus t + q_Y\cos\omega_\oplus t \right]\,.
	\label{eq:Btimedomain}
\end{equation}
The acquisition should preserve the signed source referenced quadratures, namely the in-phase and quadrature amplitudes defined relative to the measured source phase, so that the signs and correlations of $q_X$ and $q_Y$ can be determined. The same source phase convention and filtering must be applied to the data and signal templates. We distribute $11\,\mathrm{h}$ of live time over $72\,\mathrm{h}$ of elapsed time in 12 source referenced blocks, each lasting $55\,\mathrm{min}$ and centered at $t_b=(3+6b)\,\mathrm{h}$ for $b=0,\ldots,11$. This illustrative schedule samples the sidereal quadratures well; an actual acquisition uses the measured timestamps, including gaps. Broad phase coverage controls component covariance, but a three day span alone does not establish rejection of solar day backgrounds.

For a fixed mediator mass, the calibrated data can be written as
\begin{equation}
	\mathbf d=\mathbf X\mathbf q+\mathbf Z\bm\beta+\mathbf n\,,
	\label{eq:datamodel}
\end{equation}
where $\mathbf q=(q_X,q_Y)^T$ contains the two mixed couplings relevant for the measurement. The term $\mathbf Z\bm\beta$ describes backgrounds correlated with the source modulation, and $\mathbf n$ is the detector noise. Offsets and drifts can be added by augmenting $\mathbf Z$. The signal matrix $\mathbf X$ is obtained from Eq.~\eqref{eq:Btimedomain} using the measured timestamps and source reference. For zero mean Gaussian detector noise with covariance $\mathbf C=\langle\mathbf n\mathbf n^T\rangle$ and a Gaussian prior on the source correlated background amplitudes with mean $\bar{\bm\beta}$ and covariance $\mathbf P_\beta$, the residual $\mathbf r=\mathbf d-\mathbf X\mathbf q-\mathbf Z\bm\beta$ gives
\begin{equation}
	-2\ln\mathcal L = \mathbf r^T\mathbf C^{-1}\mathbf r+(\bm\beta-\bar{\bm\beta})^T\mathbf P_\beta^{-1}(\bm\beta-\bar{\bm\beta})+\mathrm{const}\,.
	\label{eq:timelikelihood}
\end{equation}
This Gaussian likelihood and nuisance parameters follow standard likelihood based inference methods~\cite{Cowan:1998ji, Cowan:2010js}. Measured calibration and source phase parameters entering $\mathbf X$ require their own constraints, while $\mathbf C$ can be obtained from the measured noise spectrum or a time series. Signal injections are needed to validate signal recovery, phase errors, gain variations, and backgrounds.

The numerical forecast reduces each observing block to one in-phase and one quadrature component. With the complex response $\widetilde{\mathcal R}_1$ defined in Eq.~\eqref{eq:G1complex}, the signal templates are
\begin{align}
	X_{b,I,X}&=-{\rm Re}\widetilde{\mathcal R}_1\,\sin\omega_\oplus t_b,\quad X_{b,Q,X}=-{\rm Im}\widetilde{\mathcal R}_1\,\sin\omega_\oplus t_b,\\
	X_{b,I,Y}&={\rm Re}\widetilde{\mathcal R}_1\,\cos\omega_\oplus t_b,\quad X_{b,Q,Y}={\rm Im}\widetilde{\mathcal R}_1\,\cos\omega_\oplus t_b.
	\label{eq:block_design}
\end{align}
This is the standard form of synchronous detection~\cite{10.1063/5.0145566}. The source modulation is encoded in $\widetilde{\mathcal R}_1$, while the slower sidereal evolution separates $q_X$ and $q_Y$ across the observing blocks. For a one sided amplitude spectral density $S_B^{1/2}$ of the calibrated magnetization equivalent channel $\mu_0M$, the peak amplitude estimator has variance $S_B/T_b$ over time $T_b$, when assuming independent white quadrature noise and fix the geometry, gain, and source phase.

After profiling the nuisance amplitudes, the Fisher matrix for $(q_X,q_Y)$ is
\begin{equation}
	\mathbf F_{XY} = \mathbf X^T\mathbf C^{-1}\!\left[\mathbf X \!-\! \mathbf Z\,(\mathbf Z^T\mathbf C^{-1}\mathbf Z+\mathbf P_\beta^{-1})^{-1}\mathbf Z^T\mathbf C^{-1}\mathbf X\right].
	\label{eq:profiled_fisher}
\end{equation}
The benchmark sets $\mathbf P_\beta^{-1}=0$, corresponding to unconstrained constant source correlated nuisance amplitudes. For fixed noise, cadence, and detector orientation, $\mathbf F_{XY}=|\widetilde{\mathcal R}_1|^2\mathbf K$, where $\mathbf K$ depends only on the observing times, sidereal templates, and noise model. At each fixed mediator mass, the projected individual $95\%$ interval half widths from the simultaneous $(q_X,q_Y)$ fit are
\begin{equation}
	|q_A|_{95} = 1.96\,\sqrt{(\mathbf F_{XY}^{-1})_{AA}}\,.
	\label{eq:marginal_intervals}
\end{equation}
The covariance between $q_X$ and $q_Y$ is retained while profiling the nuisance amplitudes. The individual intervals correspond to the one parameter $95\%$ criterion $\Delta\chi^2=3.84$, distinct from the joint $95\%$ region in the $(q_X,q_Y)$ plane, for which $\Delta\chi^2=5.99$. The full construction is given in the Supplemental Material Sec.~\ref{app:likelihood}.

The assumption of an independent white quadrature noise between observing blocks can be violated in a real measurement by phase drifts and low-frequency correlated noise, which should then be included as non-diagonal covariances or additional nuisances. The solar and sidereal daily frequencies differ by only $\Delta f\simeq3.17\times10^{-8}\,\mathrm{Hz}$, well below the $1/(72\,\mathrm{h})\simeq3.86\,\mu\mathrm{Hz}$ Fourier scale of the illustrative campaign. For this, we do not rely on separation into FFT bins and the solar and sidereal components can be fitted directly with digital lock-in or equivalent matched filter templates. Their strong covariance over $72\,\mathrm{h}$ is quantified in the Supplemental Material Sec.~\ref{app:likelihood}, motivating a longer campaign.

The east pointing benchmark is insensitive to $q_Z$. More generally, for a fixed laboratory readout direction the projection of the SCF $Z$ axis is constant in sidereal time, so the corresponding $q_Z$ signal has the same source modulation as a constant source correlated background. If that background amplitude is unconstrained, the two contributions are exactly degenerate and cannot be separated by increasing the integration time alone.

\textbf{\textit{Sensitivity forecast. ---}} 
For nearly uniform sidereal coverage, Eq.~\eqref{eq:marginal_intervals} reduces to
\begin{equation}
	|q_A|_{95}(m_\phi)\simeq\frac{1.96\,\sqrt{2}\,S_B^{1/2}}{\sqrt{T}\,\mathcal R_1(m_\phi)}\,,\qquad A=X,Y\,,
	\label{eq:qforecastsimple}
\end{equation}
where $T=\sum_b T_b$ is the total live time. For the benchmark geometry, the low-mass response is $\mathcal R_1(0)\simeq9.25\times10^{23},\mathrm{T}$. We adopt a target magnetization equivalent noise of $S_B^{1/2}=10\,\mathrm{aT}/\sqrt{\mathrm{Hz}}$, motivated by the GSO noise contribution of approximately $12,\mathrm{aT}/\sqrt{\mathrm{Hz}}$ estimated in QUAX22~\cite{Crescini:2020ykp}. With this target and the benchmark observing schedule, the projected individual $95\%$ intervals reach $|q_X|_{95}\simeq|q_Y|_{95}\simeq1.5\times10^{-43}$ at low mediator masses. By comparison, rescaling Eq.~\eqref{eq:qforecastsimple} to the $53,\mathrm{aT}/\sqrt{\mathrm{Hz}}$ noise measured by QUAX22 near the source modulation frequency gives approximately $8.0\times10^{-43}$ for the same geometry and observing schedule. While the former is a target sensitivity, the latter provides a conservative reference based on demonstrated noise performance. Both estimates assume a fixed calibrated response, independent white quadrature noise, and constant source correlated nuisance amplitudes.

\begin{figure}[htbp]
	\centering
	\includegraphics[width=\linewidth]{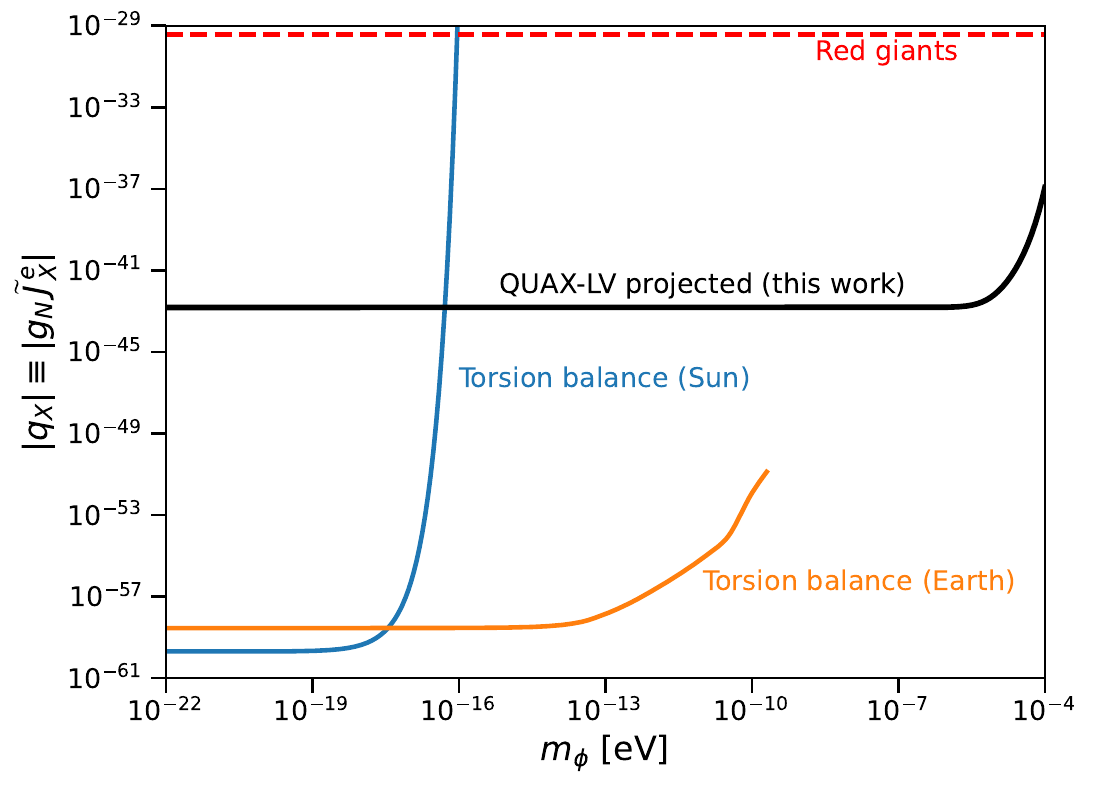}
	\caption{Projected individual $95\%$ interval for $|q_X| \equiv |g_N\,\widetilde J^e_X|$, compared with existing torsion balance limits using the Sun and Earth as sources at the $1\sigma$ level and the red-giant bound of Ref.~\cite{Carenza:2025jwn}, assuming $g_N=g_e$. The QUAX-LV target projection uses the east pointing geometry in Table~\ref{tab:optimized_geometry}, $11\,\mathrm{h}$ of live time distributed over $72\,\mathrm{h}$ of elapsed time, and a noise level of $10\,\mathrm{aT}/\sqrt{\mathrm{Hz}}$. The projected $q_Y$ intervals differ by less than $1\%$. Calibration uncertainties and time-dependent source backgrounds are not included in this forecast.}
	\label{fig:projection}
\end{figure}

Figure~\ref{fig:projection} shows the projected sensitivity compared with existing constraints. In the long-range regime, $m_\phi r\ll1$, the finite volume response approaches a constant, yielding a sensitivity plateau at $|q_A|_{95}\simeq1.5\times10^{-43}$. As the Compton wavelength $\lambda_\phi=1/m_\phi$ approaches the separation between the source and the detector, the Yukawa suppression weakens the sensitivity. For the benchmark distance $d=2\,\mathrm{cm}$, corresponding to $1/d\simeq10\,\mu\mathrm{eV}$, the projected interval reaches approximately $5.1\times10^{-42}$ at $m_\phi=20\,\mu\mathrm{eV}$. The numerical response becomes more sensitive to the sampling of nearby source and detector volume elements in this regime.

The controlled laboratory source probes separations at the centimeter scale, complementing searches employing the Earth or Sun as macroscopic sources, which provide stringent constraints in the long-range regime~\cite{Heckel:2008hw, Brown:2010dt, Smiciklas:2011xq, Carenza:2025jwn}. Red giant bounds constrain Lorentz violating electron couplings through scalar emission in electron-ion bremsstrahlung and cannot be interpreted directly as bounds on the mixed coupling $q_A = g_N\widetilde J^e_A$ without an additional assumption relating the electron and nucleon couplings~\cite{Carenza:2025jwn}.

\textbf{\textit{Conclusions. ---}} 
We have investigated the potential of QUAX to probe scalar-mediated Lorentz-violating spin-dependent interactions through a controlled laboratory source. The proposed QUAX-LV configuration combines the periodic modulation of an unpolarized Pb source with the sidereal evolution of the electron spin detector response, providing a distinctive directional signature of Lorentz violation. For an east pointing sensitivity axis, the signed source referenced quadratures allow the simultaneous extraction of the equatorial mixed couplings $q_X=g_N\widetilde J_X^e$ and $q_Y=g_N\widetilde J_Y^e$. With 11\,h of live time distributed over 72\,h and a target magnetization equivalent noise of $10\,\mathrm{aT}/\sqrt{\mathrm{Hz}}$, the benchmark configuration yields projected individual 95\% intervals of approximately $1.5\times10^{-43}$ at low mediator masses, remaining at the $10^{-43}$ level up to masses of a few $\mu\mathrm{eV}$.

This establishes QUAX as a controlled local source probe of the mixed coupling $|g_N\widetilde{\mathbf J}^{e}|$ over interaction ranges set by the centimeter-scale apparatus. The measurement complements searches using astronomical or terrestrial sources, since the source geometry, modulation, and orientation can be controlled experimentally. A realized search requires validation of the apparatus geometry, pickup response, transfer function, and noise budget. The comparison with QUAX22 in the Supplemental Material Sec.~\ref{app:geometry} supports the calculated interaction range dependence but does not establish its absolute normalization. Source-correlated backgrounds, particularly the strong solar-sidereal covariance over the illustrative three-day campaign, require auxiliary measurements or longer observations. With these experimental inputs, QUAX-LV could provide directional constraints on $q_X$ and $q_Y$, with access to $q_Z$ requiring additional orientation-dependent measurements and background discrimination.

\vspace{-0.5cm}
\section*{Acknowledgements}
This work is supported by the Istituto Nazionale di Fisica Nucleare (INFN) through the Commissione Scientifica Nazionale 2 (CSN2) Esperimento QUAX and Commissione Scientifica Nazionale~4 (CSN4) Iniziativa Specifica ``Quantum Universe'' (QGSKY). TKP is supported by the Science and Technology Facilities Council under Grant No. ST/X003167/1. 
\vspace{-0.5cm}

\bibliographystyle{apsrev4-1}
\bibliography{references}

\clearpage
\onecolumngrid
\begin{center}
\textbf{\large A proposed QUAX search for Lorentz violating scalar interactions}\\[0.08in]
{\it\large Supplemental Material}\\[0.05in]
{Domenico D'Agostino, Giovanni Carugno, Gerardo Iannone, Gaetano Lambiase, Antonello Ortolan, \newline Tanmay Kumar Poddar, Giuseppe Ruoso, Luca Visinelli}
\end{center}
\setcounter{secnumdepth}{2}
\setcounter{section}{0}
\setcounter{equation}{0}
\setcounter{figure}{0}
\setcounter{table}{0}
\renewcommand{\thesection}{S\arabic{section}}
\renewcommand{\thesubsection}{S\arabic{section}.\arabic{subsection}}
\renewcommand{\theequation}{S\arabic{equation}}
\renewcommand{\thefigure}{S\arabic{figure}}
\renewcommand{\thetable}{S\arabic{table}}
\renewcommand{\theHsection}{supp.\arabic{section}}
\renewcommand{\theHsubsection}{supp.\arabic{section}.\arabic{subsection}}
\renewcommand{\theHequation}{supp.\arabic{equation}}
\renewcommand{\theHfigure}{supp.\arabic{figure}}
\renewcommand{\theHtable}{supp.\arabic{table}}
\makeatletter
\renewcommand{\p@subsection}{}
\makeatother

\section{Finite source geometry and numerical validation}
\label{app:geometry}

The numerical calculation represents the source as 24 finite lead cylinders arranged on two concentric tracks. Their centers are placed at $\mathbf{r}_a=\left(R_a\cos\alpha_a,R_a\sin\alpha_a,0\right)$, where $a=1,\ldots,24$, $R_a=R_{\rm in}$ for $a=1,\ldots,12$ and $R_a=R_{\rm out}$ for $a=13,\ldots,24$. The corresponding azimuths are
\begin{equation}
    \alpha_a=
    \begin{cases}
    2\pi(a-1)/12, & a=1,\ldots,12,\\
    2\pi(a-13)/12+\Delta\Theta, & a=13,\ldots,24.
    \end{cases}
\end{equation}
The cylinder axes are vertical, with height $h_{\rm Pb}=60\,\mathrm{mm}$, radius $r_{\rm Pb}=25\,\mathrm{mm}$ and number density $n_N=6.8\times10^{24}\,\mathrm{cm}^{-3}$. The two GSO crystals are centered at $(R_{\rm in},0,h_{\rm Pb}/2+d)$ and $(R_{\rm out},0,h_{\rm Pb}/2+d)$ in wheel coordinates and are represented as rectangular volumes with dimensions $(25,25,11)\,\mathrm{mm}$ along the radial, tangential, and vertical directions. The wheel coordinates determine the source and detector positions, while the magnetic sensitivity direction is specified independently and is fixed to local east for the benchmark calculation. These geometrical assumptions define the numerical model used throughout the scan and forecast and are based on the available description of the experimental apparatus.

Seven quasi random coordinates are used for each sample: one selects one of the 24 source cylinders, three specify a point inside that cylinder, and three specify a point inside the detector crystal. Points are sampled uniformly within the cylindrical source volumes and the rectangular GSO volume. The sampled kernel is averaged over the joint distribution and multiplied by the total number of source nucleons. Sampling the source and detector coordinates within the same low discrepancy sequence avoids the artificial correlations that can arise when independently generated source and detector sequences are paired sample by sample.

The benchmark calculation uses $2^{16}$ joint samples per crystal and 144 source phase points, with scrambling seed 24680 for one crystal and an offset of 104729 for the second. The twelvefold symmetry of each source track permits the response to be calculated over one source modulation period rather than one complete wheel revolution. If $\psi$ denotes the phase within one source period, the physical wheel angle is $\Theta_w=\psi/12$. At every source phase the finite volume responses $\mathcal G^{(1)}$ and $\mathcal G^{(2)}$ are calculated separately and then combined with the equal weighting adopted for the common SQUID readout $\mathcal G(\Theta_w)$. The mean over the source period is removed and the complex first harmonic is extracted using the convention of Eq.~\eqref{eq:G1complex}. The complex response is retained throughout the likelihood calculation so that both its amplitude and its phase relative to the measured source reference enter the in phase and quadrature templates.

The geometry optimization and the sensitivity forecast use the same response engine. During the geometry scan, the response is evaluated at $m_\phi=(0.1,0.3,1,3,5,10,20)\,\mu\mathrm{eV}$ for every trial geometry. The two crystal responses are first combined into the single SQUID response, after which the source referenced block likelihood is constructed. The scan criterion is the mean of $\log\det F_{XY}$ over these mediator masses after profiling the constant complex source synchronous background. Once the benchmark geometry in Table~\ref{tab:optimized_geometry} is selected, the mass dependence in Fig.~\ref{fig:projection} is computed over $m_\phi\in[10^{-2},10^2]\,\mu\mathrm{eV}$ using the same $2^{16}$ samples, 144 source phase points, geometry, and combined readout prescription. Below $10^{-2}\,\mu\mathrm{eV}$, the QUAX-LV curve is extended with the constant $m_\phi r\ll1$ response. Numerical checks, including reducing the phase grid from 144 to 96 points and changing the scrambled Sobol realization at fixed $2^{16}$ samples, modify the complex response only at the percent level~\cite{Sobol:1967}.

A separate cross-check was performed by applying the same finite geometry machinery to the monopole-dipole interaction searched for in QUAX22. This comparison is not used to normalize the Lorentz violating signal. Its purpose is to test whether the numerical source and detector integration reproduces the interaction range dependence of a published QUAX-$g_pg_s$ result when the appropriate interaction kernel is inserted. For this calculation the scalar Yukawa kernel used in the Lorentz violating response is replaced by the standard derivative monopole-dipole kernel~\cite{Moody:1984ba, Dobrescu:2006au}. The same finite source integration, detector volume average, source phase sampling, and first harmonic extraction are then repeated. The reconstructed first harmonic magnetization response per unit coupling $g_p^e g_s^N$ is then $\mathcal R_{1,\rm rec}^{\rm MD}(\lambda_\phi)$.

For this historical comparison we use a reconstruction based on the public description of QUAX22, assuming source radii $0.400$ and $0.475\,\mathrm{m}$ and zero relative track phase. The response implied by the published result can be inferred independently from the reported magnetization equivalent scale and coupling limit,
\begin{equation}
    \mathcal R_{1,\rm pub}^{\rm MD}(\lambda_\phi)\equiv\frac{[\mu_0M]_{95}}{[g_p^e g_s^N]_{95}(\lambda_\phi)}\,,
    \label{eq:RpubMD}
\end{equation}
where $[g_p^e g_s^N]_{95}(\lambda_\phi)$ is the $95\%$ confidence level upper bound digitized from Fig.~4 of Ref.~\cite{Crescini:2020ykp} and $[\mu_0M]_{95}=0.73\,\mathrm{aT}$. We compare the published and reconstructed responses through
\begin{equation}
    \kappa_{2022}(\lambda_\phi) \equiv \frac{\mathcal R_{1,\rm pub}^{\rm MD}(\lambda_\phi)}{\mathcal R_{1,\rm rec}^{\rm MD}(\lambda_\phi)}\,.
    \label{eq:kappa2022}
\end{equation}
This ratio is approximately constant, with $\kappa_{2022}=4.24$ over the mass range considered in the scan and supports the reconstructed interaction range dependence over the tested interval. In the laboratory, the residual normalization difference can arise from further quantities such as the exact source and crystal positions, and the magnetic flux transfer from the GSO volumes to the SQUID, which affect the monopole-dipole and Lorentz violating responses differently. For this reason, our forecast uses directly the finite volume scalar response calculated for the benchmark geometry, with no factor $\kappa_{2022}$. A calibrated experimental prediction will require the measured apparatus geometry, transfer function, and uncertainties.

\section{Forecast likelihood and interval construction}
\label{app:likelihood}

We summarize here the statistical construction used for the projected intervals. For the east pointing configuration, the parameters of interest are $\mathbf q=(q_X,q_Y)^T$ and the mixed coupling $q_A\equiv g_N\widetilde J^e_A$. The forecast uses $N_b=12$ observing blocks. Each block is compressed into one signed in-phase component and one signed quadrature component at the source modulation frequency, giving the data vector
\begin{equation}
	\mathbf d=
	(d_{1,I},d_{1,Q},\ldots,d_{N_b,I},d_{N_b,Q})^T.
	\label{eq:app_datavector}
\end{equation}
The rapid source modulation is contained in the complex first harmonic response $\widetilde{\mathcal R}_1$ defined in Eq.~\eqref{eq:G1complex}, while the sidereal dependence follows from Eq.~\eqref{eq:Btimedomain}. The corresponding signal templates evaluated at the center $t_b$ of each observing block are
\begin{align}
	X_{b,I,X} &= -{\rm Re}\widetilde{\mathcal R}_1 \sin\omega_\oplus t_b\,, &
    X_{b,I,Y} &=  {\rm Re}\widetilde{\mathcal R}_1 \cos\omega_\oplus t_b\,,\nonumber\\
	X_{b,Q,X} &= -{\rm Im}\widetilde{\mathcal R}_1 \sin\omega_\oplus t_b\,, &
	X_{b,Q,Y} &=  {\rm Im}\widetilde{\mathcal R}_1 \cos\omega_\oplus t_b\,.
	\label{eq:app_templates}
\end{align}
The resulting signal vector is $\mathbf s = \mathbf X\mathbf q$. The opposite signs of the $X$ and $Y$ sidereal templates follow directly from $\widetilde J^e_{{\rm lab},y} = -\widetilde J^e_X\sin\omega_\oplus t +\widetilde J^e_Y\cos\omega_\oplus t$.

For a one sided magnetization equivalent noise power spectral density $S_B$, the in-phase and quadrature peak-amplitude estimators from a block of duration $T_b$ are assigned the variance
\begin{equation}
	\sigma_b^2=\frac{S_B}{T_b}.
	\label{eq:app_variance}
\end{equation}
This normalization follows, for example, for an in-phase estimator
\begin{equation}
	\widehat A_I=
	\frac{2}{T_b}
	\int_{T_b}{\rm d}t\,
	n(t)\cos\omega_s t
\end{equation}
when $S_B$ is the one sided white noise spectral density and the block contains many source modulation cycles. The quadrature estimator has the same variance. Assuming independent noise between blocks and quadratures, the covariance matrix is
\begin{equation}
	\mathbf C= {\rm diag} \left(\sigma_1^2,\sigma_1^2,\ldots, \sigma_{N_b}^2,\sigma_{N_b}^2\right)\,.
	\label{eq:app_covariance}
\end{equation}

The baseline forecast also allows for one constant source correlated amplitude in each source referenced quadrature. Their design matrix $\mathbf Z_0$ is defined by
\begin{equation}
	(\mathbf Z_0)_{b,I}=(1,0),
	\qquad
	(\mathbf Z_0)_{b,Q}=(0,1),
	\label{eq:app_Z0}
\end{equation}
for every observing block $b$. These two nuisance amplitudes are left unconstrained in the benchmark. More generally, for a nuisance design matrix $\mathbf Z$ and a Gaussian prior covariance $\mathbf P_\beta$, the Fisher matrix of the joint parameter vector $(\mathbf q,\bm\beta)$ is
\begin{equation}
	\mathbf F_{\rm full} =
    \begin{pmatrix}
	\mathbf X^T\mathbf C^{-1}\mathbf X & \mathbf X^T\mathbf C^{-1}\mathbf Z\\
	\mathbf Z^T\mathbf C^{-1}\mathbf X & \mathbf Z^T\mathbf C^{-1}\mathbf Z +\mathbf P_\beta^{-1}
	\end{pmatrix}.
	\label{eq:app_fullfisher}
\end{equation}
Profiling the nuisance amplitudes gives the Schur complement in Eq.~\eqref{eq:profiled_fisher}. The baseline calculation uses $\mathbf Z=\mathbf Z_0$ and $\mathbf P_\beta^{-1}=0$. Additional monitored environmental templates can be incorporated by augmenting the nuisance design matrix. For the linear Gaussian model used here, profiling the Gaussian nuisance amplitudes gives the same covariance for $\mathbf q$ as Gaussian marginalization.

The covariance matrix of the two inferred couplings is $\boldsymbol{\Sigma}_q = \mathbf F_{XY}^{-1}$, and the diagonal entries are the marginalized variances,
\begin{equation}
	\sigma_{q_X}^2= (\boldsymbol{\Sigma}_q)_{XX}, \qquad \sigma_{q_Y}^2= (\boldsymbol{\Sigma}_q)_{YY},
\end{equation}
and the projected individual two sided $95\%$ interval half widths are
\begin{equation}
	|q_A|_{95} = 1.96 \sqrt{(\mathbf F_{XY}^{-1})_{AA}}, \qquad A=X,Y.
	\label{eq:app_intervals}
\end{equation}
Both $q_X$ and $q_Y$ and the nuisance amplitudes are fitted simultaneously, so the covariance between the two signal components is retained. Equation~\eqref{eq:app_intervals} gives one dimensional marginalized intervals and should be distinguished from the joint $95\%$ confidence region, which for the Gaussian two-parameter problem is defined by
\begin{equation}
	\Delta\mathbf q^T \mathbf F_{XY} \Delta\mathbf q \leq 5.99.
	\label{eq:app_jointellipse}
\end{equation}

The signal templates in Eq.~\eqref{eq:app_templates} are evaluated at the center $t_b$ of each observing block. Their exact average over a block of duration $T_b$ is obtained from
\begin{align}
	\frac{1}{T_b} \int_{t_b-T_b/2}^{t_b+T_b/2} {\rm d}t\,\sin\omega_\oplus t &= \sin\omega_\oplus t_b\, {\rm sinc}\left(\frac{\omega_\oplus T_b}{2}\right),\\
	\frac{1}{T_b} \int_{t_b-T_b/2}^{t_b+T_b/2}{\rm d}t\,\cos\omega_\oplus t &= \cos\omega_\oplus t_b\, {\rm sinc}\left(\frac{\omega_\oplus T_b}{2}\right)\,.
\end{align}
For $T_b=55\,\mathrm{min}$ this factor is $0.99759$, corresponding to an increase of approximately $0.24\%$ in the projected interval widths. The block-center approximation is used in the quoted benchmark.

We also test the ability of the $72\,\mathrm{h}$ observing schedule to distinguish a sidereal signal from an unconstrained solar-day variation. In addition to the two constant source correlated nuisance amplitudes, we introduce independent sine and cosine templates at
\begin{equation}
	\omega_\odot=\frac{2\pi}{24\,\mathrm{h}}
\end{equation}
in each source referenced quadrature. The four additional nuisance columns are defined by
\begin{align}
	(\mathbf Z_\odot)_{b,I} &= \left( \sin\omega_\odot t_b,\, \cos\omega_\odot t_b,\, 0,\, 0 \right), \nonumber\\
	(\mathbf Z_\odot)_{b,Q} &= \left( 0,\, 0,\, \sin\omega_\odot t_b,\, \cos\omega_\odot t_b\right).
	\label{eq:app_solarZ}
\end{align}
The stress-test nuisance matrix is
\begin{equation}
	\mathbf Z= \left(\mathbf Z_0\;\; \mathbf Z_\odot\right),
\end{equation}
with all six amplitudes left unconstrained. The profiled Fisher matrix is recomputed using Eq.~\eqref{eq:profiled_fisher}. The degradation of each individual interval relative to the baseline is
\begin{equation}
	D_A = \sqrt{\frac{(\mathbf F_{XY,\odot}^{-1})_{AA}}{(\mathbf F_{XY,0}^{-1})_{AA}}},\qquad A=X,Y,
	\label{eq:app_degradation}
\end{equation}
where $\mathbf F_{XY,0}$ and $\mathbf F_{XY,\odot}$ denote the baseline and solar-stress-test Fisher matrices, respectively. For the stated $72\,\mathrm{h}$ schedule, the low mass calculation gives $D_X=67.8$ and $D_Y=71.5$. The large degradation follows from the close solar and sidereal frequencies over the three day elapsed span: the corresponding sine and cosine templates are nearly linearly dependent over the available observing times. The stress test is not included in the benchmark sensitivity curve; it quantifies the level of additional background discrimination required for a realized measurement.

\end{document}